\documentclass[5p,sort&compress]{elsarticle}
\usepackage{amssymb}
\usepackage{amsmath,ragged2e}
\usepackage{color,appendix}
\usepackage{graphicx}
\usepackage{caption}
\usepackage{subcaption}
\usepackage[english]{babel}
\usepackage{bm}
\usepackage[percent]{overpic}
\usepackage{multirow,rotating}
\usepackage[english]{babel}

\newcommand{\er}{$\pm$}

\newcommand{\beq}{\begin{eqnarray}}
\newcommand{\eeq}{\end{eqnarray}}
\newcommand{\be}{\begin{eqnarray}}
\newcommand{\ee}{\end{eqnarray}}
\newcommand{\bea}{\begin{eqnarray}}
\newcommand{\eea}{\end{eqnarray}}
\newcommand{\bc}{\begin{center}}
\newcommand{\ec}{\end{center}}

\newcommand{\nn}{\nonumber\\}

\definecolor{orange}{rgb}{1,.5,.3}
\definecolor{lightyellow}{cmyk}{0,0,0.5,0}
\definecolor{lightred}{rgb}{1,0.5,0.5}
\definecolor{lightgreen}{rgb}{0.8,1.0,0.8}
\definecolor{lightblue}{rgb}{0.5,0.5,1}
\definecolor{darkred}{rgb}{0.8,0,0}
\definecolor{darkgreen}{rgb}{0,0.4,0}
\definecolor{darkcyan}{cmyk}{1,0.3,0.3,0.3}
\definecolor{darkblue}{rgb}{0,0,0.6}
\definecolor{lightbrown}{rgb}{0.7,0.3,0.3}
\definecolor{darkbrown}{rgb}{0.5,0,0}
\definecolor{bluegreen}{rgb}{0,0.5,0.5}
\definecolor{mygreen}{rgb}{0.0,0.5,0.0}
\definecolor{myblue}{rgb}{0.0,0.0,0.65}
\definecolor{myblueh}{rgb}{0.0,0.0,0.65}
\definecolor{myred}{rgb}{0.72,0.0,0.}
\definecolor{mycyan}{rgb}{0.0,0.6,0.6}
\definecolor{mylila}{rgb}{0.6,0.,0.6}
\definecolor{mylila2}{rgb}{0.6,0.,0.5}
\definecolor{mygray}{rgb}{0.37,0.37,0.37}
\definecolor{myyellow2}{rgb}{0.8,0.5,0.1}
\definecolor{myyellow2b}{rgb}{0.9,0.9,0.1}
\definecolor{myyellow}{rgb}{0.6,0.6,0.}
\definecolor{mytuerkis}{rgb}{0.0,0.6,0.85}
\definecolor{grey0}{rgb}{0.7,0.7,0.7}
\definecolor{grey1}{rgb}{0.6,0.6,0.6}
\definecolor{grey2}{rgb}{0.4,0.4,0.4}
\definecolor{grey3}{rgb}{0.2,0.2,0.2}

\begin{document}
\begin{frontmatter}

\bibliographystyle{try}
\topmargin 0.1cm

\title{Singlet-octet-glueball mixing of scalar mesons }

\author[label1]{E. Klempt}
\author[label1,label2]{A.V. Sarantsev}

\address[label1]{Helmholtz--Institut f\"ur Strahlen-- und Kernphysik der Universit\"at Bonn, Nussallee 14-16, 53115 Bonn, Germany}
\address[label2]{NRC ``Kurchatov Institute'', PNPI, Gatchina 188300, Russia}

\date{\today}
\begin{abstract}
The mixing angles between scalar isoscalar resonances and a scalar glueball
are determined from their decays into two pseudoscalar mesons. For $f_0(1370)$ and
$f_0(1500)$, at most a small glueball component is admitted by the data.
The decay modes of $f_0(1710)$, $f_0(1770)$, $f_0(2020)$,
and $f_0(2100)$ require significant glueball fractions.  Above this mass, the errors
in the decay frequencies become too large to extract a glueball component. The summation 
of all observed
glueball fractions up to 2100\,MeV yields (78\er18)\%. The glueball fractions as function
of the mass are consistent with a scalar glueball at 1865\,MeV and a width of 370\,MeV
as suggested by a measurement of the yield of scalar isoscalar mesons in radiative
$J/\psi$ decays. 
\end{abstract}


\end{frontmatter}

\section{Introduction}
\bibliographystyle{try}
Light mesons with identical spin and parity are observed in meson nonets that
can be decomposed into a singlet  and an octet~\cite{GellMann:1962xb}.
Well known are the nonets of pseudoscalar mesons housing four kaons, three pions,  the $\eta$
and the $\eta'$, the vector
mesons $K^*(892), \rho(770), \phi(1020)$ and $\omega(782)$, 
and the tensor mesons $K^*_2(1430), a_2(1320)$,
$f_2^\prime(1525)$, and $f_2(1270)$. The two isoscalar states in each nonet can mix; the mixing
angle can be determined from the Gell-Mann-Okubo (GMO) mass
formula~\cite{GellMann:1962xb,Okubo:1961jc}, from the
production of mesons and from their decays: SU(3) relates the mixing
angle with the frequency of the different production and decay modes.
Mixing of the isoscalar states with heavy quarkonia is generally assumed to be negligible:
$\eta_8$ and $\eta_1$ mix but the masses of $\eta_{c}$ and $\eta_{b}$ are too large
to play a role for the low-mass mesons. Except for the pseudoscalar mesons,
most meson nonets show nearly {\it ideal} mixing: the lighter isoscalar
meson in a nonet is mainly composed of up and down quarks only while the heavier
meson can be described as mainly $s\bar s$ state.

In an analysis of radiative $J/\psi$ decays constrained by a large number of further data,
ten scalar isoscalar resonances were reported~\cite{Sarantsev:2021ein}. A  recent
analysis of the data on radiative $J/\psi$ decays - without constraints from further
data like $\pi\pi$ elastic and inelastic scattering - confirmed four of them~\cite{Rodas:2021tyb}. 
The flavor content of the two lightest isoscalar mesons
was studied by Oller~\cite{Oller:2003vf}
(see also \cite{Klempt:2021nuf}) by a fit to the two-meson residues. The $f_0(500)$
resonance was found to be $\sim$$(n\bar n+s\bar s)$
(singlet-like), the $f_0(980)$ to be $\sim$$(n\bar n+s\bar s)$ (octet-like). No gluonic contribution
was required. The interference of
the $f_0(1370)$ resonance with $f_0(1500)$ in radiative $J/\psi$ decays into $\pi\pi$ and $K\bar K$
- constructive in $\pi\pi$ and destructive in $K\bar K$ -
identified the former state as mainly singlet, the latter one as mainly octet~\cite{Sarantsev:2021ein}.
At higher masses close-by pairs of resonances were seen that were interpreted as states with octet and singlet 
$q\bar q$ components. 
Octet states should not be produced in radiative
$J/\psi$ decays. Their production was ascribed to gluonic components in their wave functions. 
The gluonic contribution to singlet states was identified by the enhanced production in 
radiative $J/\psi$ decays. The scalar glueball was
shown to extend over a wide mass range and to be part of several scalar mesons.
Its mass and width were determined to $M=(1865\pm25)$\,MeV, $\Gamma=(370^{+30}_{-20})$ MeV.
 This mass is just compatible with the result from unquenched  lattice calculations
which predict a scalar glueball mass of $(1795\pm60)$\,MeV~\cite{Gregory:2012hu}. 

In this Letter we determine the fractional glueball contents and the mixing angles for
these and higher-mass scalar mesons.
In Section~\ref{II} we give the relations used to determine
mixing angles.   In Section~\ref{IV} we determine the scalar mixing angle from
a fit to $f_0(1370)$ and $f_0(1500)$ decays, search
for a glueball fraction in their wave function, and discuss if they form,
together with $a_0(1450)$ and $K^*_0(1430)$, a valuable nonet. 
In Sections~\ref{V} - \ref{VII} we discuss mixing angles and glueball fractions
of $f_0(1710)$/$f_0(1770)$, $f_0(2020)$/$f_0(2100)$, and of 
$f_0(2200)$/$f_0(2330)$,
and possible nonet assignments.
The Letter ends with a discussion of the results and a short summary (Section~\ref{VIII}).
\section{\label{II}SU(3) relations}

\begin{figure}[h!]
\phantom{zzzzzzzzzzzzzzzzzzzzzzzzzzzzzzzzzzzzzzzzzzzzzzzzzzzzzzzzzzzzzzz}\vspace{5mm}
\includegraphics[width=0.48\textwidth]{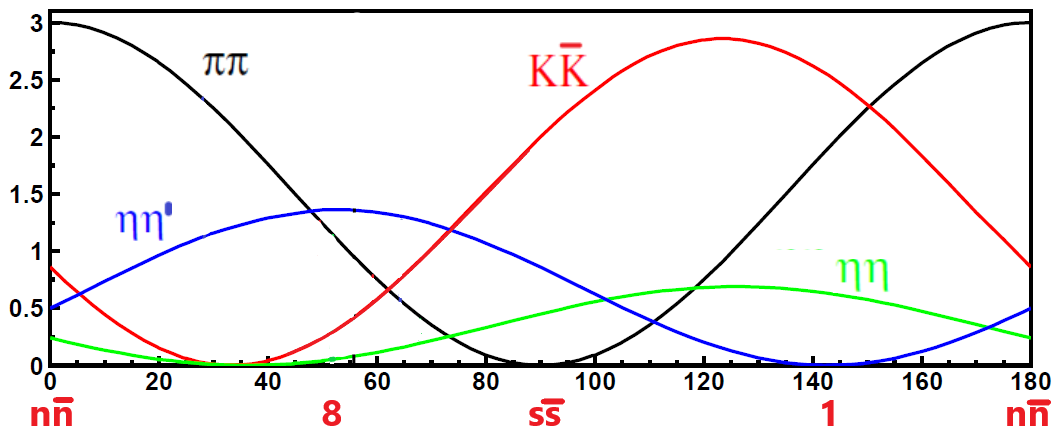}\vspace{-1mm}
\caption{\label{fig:su3}The SU(3) structure constants as functions of the mixing angle $\varphi$.
For $\varphi=0$, the meson is a $n\bar n$, for $\varphi=90^\circ$, it is a $s\bar s$
state. Singlet and octet configurations are indicated.
}\vspace{-1mm}
\end{figure}

The mixing angle of scalar mesons can be derived from their decays exploiting SU(3)
relations~\cite{Peters:1995jv,Anisovich:2011zz,Amsler:2008zza}.
A singlet isoscalar meson may decay into
$\pi\pi$, $K\bar K$, $\eta\eta$ and $\eta\eta'$ with squared
coupling constants proportional to 3\,:\,4\,:\,1\,:\,0. An octet isoscalar meson
has the corresponding squared coupling constants 3\,:\,1\,:\,1\,:\,4. These relations hold true when
$\eta$ and $\eta'$ are pure octet and singlet states. However, singlet and octet isoscalar mesons mix, 
and the ratios for decays into $\eta$ and $\eta'$ are modified due to a finite pseudoscalar mixing angle. 
Singlet and octet mixing also occurs for scalar isoscalar mesons, and the decay coupling constants
depend on the scalar mixing angle.  Further, decays into $K\bar K$ are suppressed by a suppression factor~$\lambda$.
When these complications are taken into account,
the SU(3) relations governing the decays of $q\bar q$ mesons and glueballs into two pseudoscalar
mesons can be cast into a form presented in Table~\ref{SU3} and, in graphical form, 
in Figure~\ref{fig:su3}. We call the higher-mass state
$f^H$. For the lower-mass state $f^L$, orthogonal in SU(3), $\cos\varphi^{\rm s}$ is substituted by
$\sin\varphi^{\rm s}$, and $-\sin\varphi^{\rm s}$ by $\cos\varphi^{\rm s}$.

In Table~\ref{SU3}, the scalar mixing angle $\varphi^{\rm s}$ is given in the quark basis 

\bc$
\left( \begin{array}{c}
f^H\\
f^L\\
\end{array}\right)
=
\left( \begin{array}{cc}
\cos\varphi^{\rm s}& -\sin\varphi^{\rm s}  \\
\sin\varphi^{\rm s}  & \cos\varphi^{\rm s}\\
\end{array}\right)
\left( \begin{array}{c}
|n\bar n> \\
|s\bar s>
\end{array}\right)$
\quad (1)\ec
In the singlet/octet basis, the mixing angle is given by 
\bc$\hspace{-1mm}
\left( \begin{array}{c}
f^H\\
f^L\\
\end{array}\right)
=
\left( \begin{array}{cc}
\cos\vartheta^{\rm s}& -\sin\vartheta^{\rm s}  \\
\sin\vartheta^{\rm s}  & \cos\vartheta^{\rm s}\\
\end{array}\right)
\left( \begin{array}{c}
|8> \\
|1>
\end{array}\right)$
\qquad\  (2)\ec
\noindent
where $f^H$ is the heavier isoscalar meson.
The two angles are related by $\varphi^{\rm s}=\vartheta^{\rm s}+(90-35.3)^\circ$.
$\vartheta_{\rm ideal}=35.3^\circ$ is the {\it ideal} mixing angle with
$\tan\vartheta_{\rm ideal}=1/\sqrt2$. Fig.~\ref{fig:su3} is very similar
to Fig.\,15.2 in the Review of Particle
Physics (RPP2020)~\cite{Zyla:2020zbs}.  

The pseudoscalar mixing angle in the quark basis is given by
\bc$
\left( \begin{array}{c}
\eta\\
\eta'\\
\end{array}\right)
=
\left( \begin{array}{cc}
\cos\phi^{\rm ps}& -\sin\phi^{\rm ps}  \\
\sin\phi^{\rm ps}  & \cos\phi^{\rm ps}\\
\end{array}\right)
\left( \begin{array}{c}
|n\bar n> \\
|s\bar s> \\
\end{array}\right)$
\ \ (3)\ec 
and in the singlet/octet basis by 
\bc$
\left( \begin{array}{c}
\eta\\
\eta'\\
\end{array}\right)
=
\left( \begin{array}{cc}
\cos\theta^{\rm ps}& -\sin\theta^{\rm ps}  \\
\sin\theta^{\rm ps}  & \cos\theta^{\rm ps}\\
\end{array}\right)
\left( \begin{array}{c}
|8> \\
|1> \\
\end{array}\right)$
\quad\ (4)\ec
\setcounter{equation}{4}

The octet and singlet decay constants are different, and the mixing between the
two isoscalar states is described by two mixing angles
$\theta_1^{\rm ps}$ and $\theta_8^{\rm ps}$~\cite{Feldmann:1998su}. For the
study of scalar mesons into two pseudoscalar mesons, we use the quark basis.
In this basis the two mixing angles  $\phi_1^{\rm ps}$ and $\phi_8^{\rm ps}$ are
very similar in magnitude~\cite{Chen:2014yta}; we use
$\phi^{\rm ps}=(39.3\pm1.0)^\circ$~\cite{Feldmann:1999uf}. In the singlet-octet basis,
this mixing angle corresponds to
$\theta^{\rm ps}=(39.3+35.3-90)^\circ= -(15.4\pm1.0)^\circ$.

The strangeness suppression factor $\lambda$ can be derived from $J/\psi$ decays into
baryons. The partial decay widths are taken from the RPP2020~\cite{Zyla:2020zbs}. Corrected for the phase space they
yield reduced widths $\Gamma^\prime$.
\begin{align}
\lambda&=&\Gamma^\prime _{J/\psi\to \Sigma^+\bar\Sigma^-}/\Gamma^\prime _{J/\psi\to p\bar p}&=&0.88\pm0.14\nonumber\\
\lambda&=&\Gamma^\prime _{J/\psi\to \Sigma^0\bar\Sigma^0}/\Gamma^\prime _{J/\psi\to n\bar n}&=&0.69\pm0.15\nonumber \\
\lambda&=&\Gamma^\prime _{J/\psi\to \Lambda\bar\Lambda}/\Gamma^\prime _{J/\psi\to n\bar n}&=&1.04\pm 0.17\nonumber\\
\lambda^2&=&\Gamma^\prime _{J/\psi\to \Xi^-\bar\Xi^+}/\Gamma^\prime _{J/\psi\to p\bar p}&=&(0.70\pm 0.06)^2\nonumber
\end{align}
This gives a mean value of
\begin{eqnarray}
\lambda&=&0.84\pm0.04
\end{eqnarray}

This value is consistent with $\lambda=0.77\pm0.10$ obtained from an analysis of the decay of tensor
mesons~\cite{Peters:1995jv}. The suppression factor is valid over a wide momentum range.

The strangeness suppression factor is a measure of SU(3) violation. Note that
the strangeness suppression factor in 
\begin{table}[pt]
\caption{\label{SU3}SU(3) structure constants for the decays of $(q\bar q)$-mesons,
$\gamma^\alpha _{q\bar q}$,
and glueballs,  $\gamma^\alpha_G$, into two pseudoscalar mesons given by quark
combinatorics. See text for the definition of the mixing angles. The creation of an $s\bar s$
pair is supposed to be suppressed by a factor $\lambda$. \vspace{-3mm}}
\renewcommand{\arraystretch}{1.2}
\bc
\begin{tabular}{cc}
\hline\hline
Decay  &Coupling constants $\gamma_\alpha ^q$\\ \hline
$f^H \to\pi\pi$ & $\sqrt{3}\cos\varphi^{\rm s}$ \\
$f^H  \to K\bar K$ & $(-\sqrt{2}\sin\varphi^{\rm s}+\sqrt{\lambda}\cos\varphi^{\rm s})$\\
$f^H \to \eta\eta$& $(\cos^2\phi^{\rm ps}\;\cos\varphi^{\rm s}
-\sqrt{2\lambda}\;\sin^2\phi^{\rm ps}\;\sin\varphi^{\rm s})$\\
$f^H \to \eta\eta'$ & $\frac{1}{\sqrt2}\sin 2\phi^{\rm ps}\;(\cos\varphi^{\rm s} +\sqrt{2\lambda}\;\sin\varphi^{\rm s})$\\
$f^L \to\pi\pi$ & $ \sqrt{3}\sin\varphi^{\rm s}$ \\
$f^L  \to K\bar K$ & $ (\sqrt{2}\cos\varphi^{\rm s}+\sqrt{\lambda}\sin\varphi^{\rm s})$\\
$f^L \to \eta\eta$& $ (\cos^2\phi^{\rm ps}\;\sin\varphi^{\rm s}
+\sqrt{2\lambda}\;\sin^2\phi^{\rm ps}\;\cos\varphi^{\rm s})$\\
$f^L \to \eta\eta'$ & $\frac{1}{\sqrt2}\sin 2\phi^{\rm ps}\;(\sin\varphi^{\rm s}-\sqrt{2\lambda}\;\cos\varphi^{\rm s})$\\
$a \to \eta\pi$& $\sqrt{2}\cos\phi^{\rm ps}$\\
$a \to \eta'\pi$&  $\sqrt{2}\sin\phi^{\rm ps}$\\
$a \to K\bar K$& $\sqrt{\lambda}$\\
$K\to K\pi$ & $\sqrt{3/2}$\\
$K\to K\eta$ & $\frac{1}{\sqrt2}(\cos\phi^{\rm ps} -\sqrt{2\lambda}\sin\phi^{\rm ps})$\\
$K\to K\eta'$ &  $\frac{1}{\sqrt2}(\sin\phi^{\rm ps} +\sqrt{2\lambda}\cos\phi^{\rm ps})$\\
\hline\hline
  &Coupling constants $\gamma^\alpha _G$\\ \hline
$G \to\pi\pi$ & $\sqrt{2}$ \\
$G \to K\bar K$ & $\sqrt{\frac23}(1+\sqrt\lambda)$\\
$G \to \eta\eta$& $\sqrt{\frac23}(\cos^2\phi^{\rm ps}\;+\;\sqrt{\lambda}\;\sin^2\phi^{\rm ps})$\\
$G\to \eta\eta'$ & $\frac{1}{\sqrt{3}}(1-\sqrt\lambda)\sin 2\phi^{\rm ps}$\\
\hline\hline
\end{tabular}\vspace{-10mm}
\ec
\end{table}
exclusive two-body decays is different from
the strangeness suppression in fragmentation. Here, we compare, e.g., $a_2(1320)\to K\bar K$
with $a_2(1320)\to \pi\eta$; in fragmentation one compares $a_2(1320)\to K\bar K$
with all $a_2(1320)$ decays including, e.g., $a_2(1320)\to\rho\pi$.

We write the wave function of a scalar states in the form
\beq
~\hspace{-6mm}f^{\rm nH} _0(xxx)=\left(n\bar n\cos\varphi^{\rm s} _{\rm n}-s\bar s\sin\varphi^{\rm s} _{\rm n}\right )\cos\phi^G _{\rm nH} + G\sin\phi^G _{\rm nH}\nn
~\hspace{-8mm}f^{\rm nL} _0(xxx)=\left(n\bar n\sin\varphi^{\rm s} _{\rm n}+s\bar
s\cos\varphi^{\rm s} _{\rm n}\right )\cos\phi^G _{\rm nL} + G\sin\phi^G _{\rm nL}\nonumber\eeq
$\varphi^{\rm s} _{\rm n}$ is the scalar mixing angle, $\phi^G _{\rm nH}$ and $\phi^G _{\rm nL}$ 
are the meson-glueball mixing angles of the high-mass state H and of the low-mass state L in the nth nonet.
The fractional glueball content of a meson is given by $\sin^2\phi^G _{\rm nH}$ or $\sin^2\phi^G _{\rm nL}$.

The $q\bar q$ component of a scalar meson couples to the final states with the
SU(3) structure constant $\gamma_\alpha$ and with a decay coupling constant $c_n$.
The $\gamma_\alpha$ depend on the final state, 
the constants $c_{\rm n}$
depend on the SU(3) nonet: the decays of $f_0(1500)$, $f_0(1370)$, $a_0(1450)$, and $K^*(1430)$
should be described by the constant $c_1$, $f_0(1770)$\,/\,$f_0(1710)$ require a different value $c_2$.
The SU(3) structure constants $\gamma_\alpha$ of a $q\bar q$ singlet and of a glueball  are identical.
There is one coupling constant $c_{G}$ for the glueball contents of all scalar mesons.

The coupling of a meson in nonet $n$ to the final state $\alpha$ can be written as
\beq
g_\alpha ^{\rm n} =& c_{\rm n} \gamma_\alpha ^q + c_G\gamma_\alpha ^G\,. 
\label{cc}
\eeq
The fit to the data described in Ref.~\cite{Sarantsev:2021ein} returns the squared coupling constants
$(g^{\rm n} _\alpha)^2$. From these coupling constants, the partial decay widths were derived
using the expression 
\beq
\hspace{-7mm}M \Gamma_{\alpha}^n =  \int\limits_{\rm
threshold}^{\infty}\frac {ds}{\pi} \frac{
\{(g_\alpha^n)^2\rho_{\alpha}^n(s)\}^2}{(M\,^2-s)^2+
(\sum\limits_\alpha g_{\alpha}^n\,^2\rho_{\alpha}^n(s))^2}\,,
\label{br4}
\eeq
where $s$ is the two-meson invariant mass, $\rho_{\alpha}^n(s)$ the phase space,
$M$ is the nominal meson mass.
The partial decay widths are given in Table~3 of Ref.~\cite{Sarantsev:2021ein}.
For $a_0(1450)$ and $K^*_0(1430)$ we determine the coupling constants from
their partial decay width into a final~state~$\alpha$.

\begin{figure}[b!]
\includegraphics[width=0.47\textwidth,height=0.35\textheight]{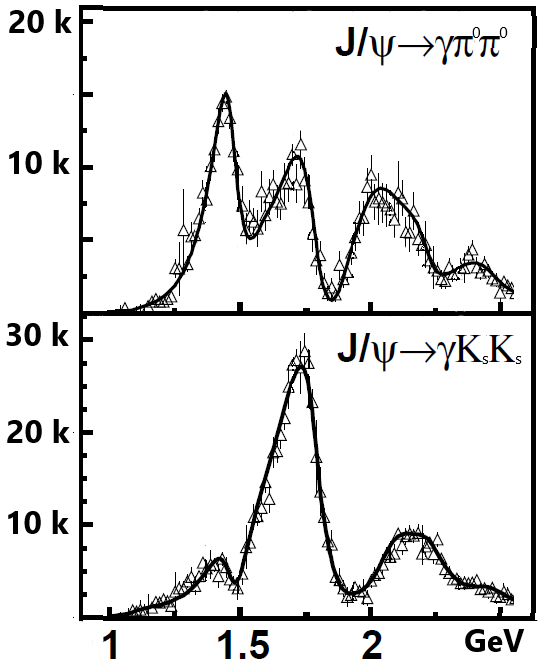}
\caption{\label{fig:data}Radiative $J/\psi$ decays into $\pi\pi$ and $K\bar K$.
The data are from Refs.~\cite{Ablikim:2015umt,Ablikim:2018izx}, the fit curve from
Ref.~\cite{Sarantsev:2021ein}.
}
\end{figure}

\section{\label{IV}$f_0(1370)$ and $f_0(1500)$: mixing angle and nonet}
Figure~\ref{fig:data} shows the $\pi\pi$ and $K\bar K$ invariant mass distributions
produced in radiative $J/\psi$ decays. In the $\pi\pi$ spectrum, an enhancement
is seen just below 1500\,MeV falling down sharply above 1500\,MeV. We assume that $f_0(1370)$ is
a mainly-singlet and $f_0(1500)$ a mainly-octet state, as suggested in Ref.~\cite{Sarantsev:2021ein}.
The two amplitudes then interfere constructively, creating the observed peak. In the
$K\bar K$ final state, the amplitudes for $f_0(1370)$ and $f_0(1500)$ interfere destructively,
there is little intensity and a clear minimum at 1500\,MeV. The constructive interference in $\pi\pi$
and the destructive interference in $K\bar K$ requires  interference between octet and singlet 
amplitudes. The fit of Rodas {\it et al.}~\cite{Rodas:2021tyb} to the data on radiative $J/\psi$ 
decay into $\pi\pi$ and $K\bar K$ (without constraints 
from $\pi\pi$ scattering) did not include $f_0(1370)$. Consequently, the minimum at 1500\,MeV
in the $K\bar K$ mass distribution was not well reproduced.

To be more quantitative, we have fit the data to determine the scalar mixing angle
and to study a possible glueball component.
Table~\ref{1500} presents the resonances in the first nonet of scalar mesons above 1\,GeV with
their decay modes, and  the fitted and the experimental squared coupling constants.
The ``experimental" squared coupling constants are taken from Ref.~\cite{Sarantsev:2021ein},
those of the isovector and isodoublet mesons were calculated from the partial and total decay-widths  given in
the RPP2020~\cite{Zyla:2020zbs}.
Only the isoscalar mesons are used for the fits described below. The coupling constants 
of isovector and isodoublet mesons are predictions.

In a first step, we assume that there is no glueball in the wave functions of
$f_0(1370)$ and $f_0(1500)$ and fit the scalar mixing angle $\varphi^{\rm s} _1$ and the
coupling $c_1$ only. Note that $c_1$ is one constant for a full nonet.
The fit to the eight branching ratios returns
\beq
\hspace{-8mm}\varphi^{\rm s} _1=(56\pm8)^\circ, &\hspace{-2mm} \vartheta^{\rm s} _1=(1\pm8)^\circ,
&\hspace{-2mm}\chi^2/N_F=19/(8-3)
\eeq
Obviously, the $f_0(1370)$ is compatible with a pure singlet, $f_0(1500)$ with a pure octet
state.
\begin{table}[pt]
\caption{\label{1500}Coupling constants of decays of mesons in the
first nonet of scalar mesons. The fit to the decays of the two isoscalar mesons yields
$c_1=0.21\pm0.02$, $c_G=0.34$, $\varphi^{\rm s} _1=(64\pm12)^\circ$, 
$\phi_{\rm 1H} ^G=(5\pm8)^\circ$,
 $\phi^G _{\rm 1L}=(10\pm6)^\circ$, $\chi^2=6.3$ for 8 data points.\vspace{-5mm}
}
\bc
\renewcommand{\arraystretch}{1.2}
\begin{tabular}{rcccc}
\hline\hline
&$g_{fit}^2$&$g_{exp}^2$\\[0.3ex]
\hline
$f_0(1500)\ \to\pi\pi$   & 0.037 & 0.034\er0.007     \\
$K\bar K$                & 0.018 & 0.014\er0.004 \\
$\eta\eta$                 & 0.004 & 0.006\er0.002  \\
$\eta\eta'$               & 0.050  & 0.061\er0.014  \\
\hline
$f_0(1370)\ \to\pi\pi$    & 0.166 & 0.226\er0.048\\
$K\bar K$                  & 0.151 & 0.116\er0.048\\
$\eta\eta$                  & 0.042 & 0.040\er0.011\\
$\eta\eta'$                 & 0.003 & 0.025\er0.019\\
\hline
$a_0(1450)\to K\bar K$ & 0.038 & 0.048\er0.016 \\
$\pi\eta$                     & 0.053 & 0.047\er0.010\\
$\pi\eta'$                 & 0.035 & 0.026\er0.013\\
\hline
$K^*_0(1430)\to K\pi$  & 0.066 & 0.450\er0.048\\
$K\eta$                       & 0.002 & 0.045\er0.020\\
$K\eta'$                      & 0.059  & -\\\hline\hline
\end{tabular}\vspace{-5mm}
\ec
\end{table}

In a next step, we allow for mixing of the two isoscalar mesons with a glueball and
impose $\vartheta^{\rm s} _1=0$. This fit returns $c_G=0.34$ which we freeze for the subsequent
fits: the glueball decay-coupling-constant is the same for all scalar mesons to which the
glueball contributes. With $c_G=0.34$ fixed, the final fit
returns mixing angles and a $\chi^2$:
\begin{align}
\hspace{-6mm}\varphi^{\rm s} _1&=(64\pm12)^\circ, & \vartheta^{\rm s} _1=(9\pm12)^\circ &\\
\phi^G _{\rm 1H}&=\ (5\pm8)^\circ, & \phi^G _{\rm 1L}=(10\pm6)^\circ, &\quad \chi^2/N_F=6.3/4\nonumber
\end{align}

This is a significant improvement, and these two mesons may contain
some glueball fraction. In any case, the glueball component given by $\sin^2\phi^G_{\rm 1H,1L}$
is small.  It is estimated to 4.0\er2.5\% for $f_0(1370)$
and to be less than 5\% in $f_0(1500)$. The coupling constant $c_1$ is determined to
\beq
c_1 = 0.21\pm 0.02 
\eeq

With the value of $c_1$ and $\varphi_1 ^{\rm s}$
as determined above we can calculate the branching ratios
expected for the scalar mesons $a_0(1450)$ and $K^*_0(1430)$. These predictions
are also given in Table~\ref{1500}.
There is one striking incompatibility: Experimentally, the $K^*_0(1430)$ resonance is nearly elastic but the
coupling constant is predicted to be rather small.
This large discrepancy is very intriguing, in particular since all other branching ratios are compatible
with the interpretation of $K^*_0(1430)$, $a_0(1450)$,
$f_0(1370)$ and $f_0(1500)$ as members of the same SU(3) nonet.
Neither the  $K^*_0$ $(1430)\to K\pi$ branching ratio nor the $a_0(1450)\to K\bar K$ or $\pi\eta$
branching ratios depend on the scalar mixing angle. Obviously, the discrepancy does not depend on
the scalar mixing angle. The two
mesons $K^*_0(1430)$ and $a_0(1450)$ are commonly be interpreted as members of the same nonet
but the $K^*_0$ $(1430)\to K\pi$ decay mode is at variance with this interpretation.
We have no explanation of this discrepancy.

The Gell-Mann--Okubo mass formula 
\be
\tan\vartheta^{\rm s} _1=\frac{4m_K-m_a-3m_{f'}}{2\sqrt2(m_a-m_K)}\nonumber
\ee
yields unstable results.
We identify $f_0(1370)=f$ (singlet) and $f_0(1500)=f'$ (octet).
The mass difference between the masses of $K^*_0(1430)$ and $a_0(1450)$
is nearly compatible with zero, and the  mixing angle is nearly unconstrained.

\section{\label{V}$f_0(1710)$ and $f_0(1770)$: mixing angle and nonet}

The two mesons $f_0(1710)$ and $f_0(1770)$ are rather close in mass. They produce a
large peak in the $K\bar K$ invariant mass distribution (see Fig.~\ref{fig:data}), with
much higher intensity than in the $\pi\pi$ spectrum.  This could indicate a large
$s\bar s$ component in the wave function in one of the two states. But the partial-wave
analysis revealed considerably larger $K\bar K$ yields than $\pi\pi$ yields for both
resonances. This is impossible for two states belonging to the same nonet. Indeed, a
fit to branching ratios of of $f_0(1770)$ and $f_0(1710)$ (see Table~\ref{1770})
with an arbitrary mixing angle but with no glueball contribution fails. Hence we
added a glueball component to the wave function and assumed a mainly-octet
$q\bar q$ structure for $f_0(1770)$, a mainly-singlet $q\bar q$
structure for $f_0(1710)$ and constructive interference between glueball and $f_0(1770)$
in the $K\bar K$ decay amplitude. These assumptions entail for $f_0(1710)\to \pi\pi$ decays a constructive
interference between the $q\bar q$ and glueball amplitudes,
a destructive interference for $f_0(1770)\to \pi\pi$, and constructive interference
between the $q\bar q$ and glueball amplitudes for both resonances in the $K\bar K$ decay mode.
This pattern can be seen in Fig.~\ref{fig:data}:
the $\pi\pi$ invariant mass distribution is rapidly falling down at 1750\,MeV while the
$K\bar K$ mass distribution exhibits a strong peak at this mass.
\begin{table}[h]
\caption{\label{1770}Coupling constants of decays of mesons in the
second nonet of scalar mesons. The fit yields  $c_2=0.38\pm0.04$,
$c^G=0.34$, $\varphi^{\rm s} _2=(41\pm4)^\circ$, $\phi_G ^{\rm 2H}=-(30\pm6)^\circ$,
$\phi_G ^{\rm 2L}=-(20\pm 5)^\circ$. The $\chi^2=44$ for 8 data points.\vspace{-4mm}
}
\bc
\renewcommand{\arraystretch}{1.2}
\begin{tabular}{rccc}
\hline\hline
 &$g_{fit}^2$& $g^2_{exp}$  \\[0.3ex]\hline
$f_0(1770)\to\pi\pi$      & 0.036 & 0.042\er 0.012  \\
$K\bar K$                     & 0.121 & 0.124\er 0.037  \\
$\eta\eta$                 & 0.010 & 0.017\er 0.004  \\
$\eta\eta'$                & 0.130 & 0.030\er 0.018  \\
\hline
$f_0(1710)\to\pi\pi$        & 0.063 & 0.090\er0.031 \\
$K\bar K$                      & 0.170 & 0.186\er0.043 \\
$\eta\eta$                     & 0.036 & 0.145\er0.051 \\
$\eta\eta'$                   & 0.007 & 0.134\er0.059 \\
\hline\hline
\end{tabular}\vspace{-3mm}
\ec
\end{table}

The $\chi^2$ of 44 for 4 degrees of freedom is unacceptably large. The large $\chi^2$ 
can be traced to stem mainly from the yields of $\eta\eta'$ decays. 
The fit assigns the $\eta\eta'$ intensity to $f_0(1710)$ that is interpreted here as 
mainly-singlet state. Little $\eta\eta'$ intensity is assigned to $f_0(1770)$. We do not exclude
that this assignment by the fit is missleading: these two resonances are very close in mass
and it could be difficult to separate contributions reliably. We emphasize that the quality
of the different data sets is rather different: the $\pi\pi$ and $K\bar K$ data are the most reliable
ones. Here, the $S$-wave has been extracted in a model-independent way. The $\eta\eta$ 
$S$-wave contribution used here stems from an energy-dependent partial-wave analysis
performed by the BESIII collaboration using Breit-Wigner representations for resonances. 
Unfortunately, the original data are not publicly available. Data on $\eta\eta'$ are poor.  

The most important information on the glueball content stems from the $\pi\pi$ and
$K\bar K$ decay modes. The fit yields\\[-4ex]
\begin{align}
\hspace{-6mm}\varphi^{\rm s} _2&=(41\pm4)^\circ, & \vartheta^{\rm s} _2=-(14\pm4)^\circ &\\
\phi^G _{\rm 2H}&=-(29\pm6)^\circ, & \phi^G _{\rm 2L}=-(20\pm5)^\circ, &\quad \chi^2/N_F=44/4\nonumber
\end{align}
The glueball content of $f_0(1710)$ is determined to $\sin^2\phi^G _{\rm 2L}$= (12\er6)\% and of 
$f_0(1770)$ to (25\er10)\%. The scalar mixing angle $\vartheta^{\rm s}_2$
is not compatible with a simple singlet-octet configuration (with $\vartheta^{\rm s} _2=0)$.

In 1650 to 1850\,MeV mass region an $a_0$ and a $K_0^{*}$ are both missing to complete
a SU(3) nonet.  Neither an $a_0$ nor a $K_0^{*}$  resonance was reported in the RPP2020
in this mass range. There are two possible interpretations.

First, the two resonances  $f_0(1710)$ and $f_0(1770)$ could only be one single state - let us call it $f_0(1750)$ - and this
could be the glueball. However, the squared masses of the five singlet
and of the five octet scalar isoscalar mesons fall onto linear $(n, M^2)$ trajectories
(see Fig.~3 in Ref.~\cite{Sarantsev:2021ein}).
Clearly, its is difficult to take out the two states at 1700\,MeV: a gap would be created.
Further, the fit to the data deteriorates significantly when one of the two
resonances is taken out.

The second possibility is that an $a_0$ and a $K_0^{*}$ do exist. Indeed, 
the BABAR collaboration reported a fit to the Dalitz plots $\eta_c\to \eta'K^+K^-$, $\eta'\pi^+\pi^-$,
$\eta\pi^+\pi^-$ and identified a new isovector state $a_0(1700)$ with $M=(1704\pm 5\pm 2)$\,MeV
and $\Gamma=(110\pm 15\pm11)$\,MeV~\cite{BaBar:2021fkz}. 
In an
analysis of the $K\pi$ S-wave, the authors of Ref.~\cite{Anisovich:1997qp} fit the
data above 900\,MeV and find a further pole slightly above
1800\,MeV and a width of 200 to 260\,MeV. The existence of a  $K_0^{*}(1800)$
is at least not ruled out. However, the mass values of
$a_0(1700)$, $K^*_0(1800)$, $f_0(1710)$, $f_0(1770)$ do not yield a 
consistent nonet: The GMO formula 
suggests $f_0(1770)$ to be a dominantly $n\bar n$ state even though it
decays preferentially into $K\bar K$. Yet, the uncertainty in the
mixing angle is substantial.

\section{\label{VI}$f_0(2020)$ and $f_0(2100)$: mixing angle and nonet}
Table~\ref{2100} presents the results of our fit for the third pair of scalar mesons
above 1\,GeV. There is strong interference
between these two resonances but also with both isoscalar mesons of the two 
neighboring nonets. Again, the couplings to $K\bar K$ of both isoscalar
resonances in this nonet are larger than the couplings to $\pi\pi$. This is not possible
without interference with another amplitude. Again, we assume the scalar glueball
to make up a fraction of the mesonic wave functions of $f_0(2020)$ and $f_0(2100)$.

\begin{table}[h]
\caption{\label{2100}Coupling constants of decays of mesons in the
third nonet of scalar mesons. The fit yields Coupling constants of $f_0(2100)$ and
$f_0(2020)$ decays.
 $c_3=0.51\pm 0.06$, $c_G=0.34$, $\phi^{\rm s} _3=51\pm4$, $\phi^G _{\rm 3H}=-23\pm7$,
 $\phi^G _{\rm 3L}=-24\pm6$. $\chi^2=15$ for 6 data points.\vspace{-3mm}
}
\bc
\renewcommand{\arraystretch}{1.2}
\begin{tabular}{rccc}
\hline\hline 
 &$g_{fit}^2$&$g_{exp}^2$ \\[0.3ex]\hline
$f_0(2100)\to\pi\pi$     & 0.104 &  0.105\er0.020    \\
$K\bar K$   & 0.211 &  0.193\er0.026 \\
$\eta\eta$   & 0.015 &  0.149\er0.039 \\
\hline 
$f_0(2020)\to\pi\pi$       & 0.192 &  0.192\er0.022\\
$K\bar K$                      & 0.276 &  0.297\er0.017\\
$\eta\eta$                   & 0.067 &  0.055\er0.010\\
\hline\hline
\end{tabular}
\ec
\vspace{-3mm}
\end{table}
The data are again fit with an unacceptable $\chi^2$ that is due to the
$f_0(2100)\to\eta\eta$ decay mode. The $\pi\pi$ and $K\bar K$ decay modes are described
excellently. Therefore we think, we can extract the glueball content reliably.
We find
\begin{align}
\hspace{-6mm}\varphi^{\rm s} _3&=\hspace{-3mm}&(51\pm4)^\circ, &\hspace{3mm} \vartheta_{\rm s}&\hspace{-3mm}=&-(4\pm4)^\circ &\\
                       \phi^G _{\rm 3H}&=\hspace{-3mm}&-(23\pm7)^\circ, & \hspace{3mm}\phi^G _{\rm 3L}&\hspace{-3mm}=&-(24\pm6)^\circ, &  \chi^2/N_F=15/3\nonumber
\end{align}
and the glueball content of $f_0(2020)$  and $f_0(2100)$ is determined to
(16\er9)\% and (17\er8)\%. The nonet mixing angle is compatible with $f_0(2020)$ 
being a singlet and $f_0(2100)$ being an octet. The $K^*_0(1950)$ and
$a_0(1950)$ could be the partners to form a full nonet. 
The Gell-Mann--Okubo mass formula does not constrain the mixing angle due to
the large uncertainties in the masses.

\section{\label{VII}$f_0(2200)$ and $f_0(2330)$}
Too little is known about these two resonances. For an octet state, a $\pi\pi$ : $K\bar K$ : $\eta\eta$
ratio of 3 : 1 : 1 is expected (for a pseudoscalar mixing angle $\vartheta_{\rm\,ps}=0$), not incompatible
with the radiative decay rates (in units of $10^{-5}$) of 4\er2 : 2.5\er0.5 : 1.5\er0.4  
observed for $J/\psi\to \gamma f_0(2330), f_0(2330)\to \pi\pi, K\bar K, \eta\eta$ \ even though
some glueball admixture is certainly not excluded. For $f_0(2200)$ as singlet, we expect these ratios to be
3 : 4 : 1 which are compatible with 5\er 2 : 5\er 2 : 1.5\er 0.4. A glueball admixture is not required and not forbidden.

\section{\label{VIII}Discussion and Summary}
The nonet containing $f_0(1370)$ and $f_0(1500)$ and the one containing 
$f_0(2020)$ and $f_0(2100)$ are shown to have mixing angles very close to 
the conjecture proposed in Ref.~\cite{Sarantsev:2021ein}: the lower-mass
states are compatible with a pure singlet $q\bar q$ component, the higher-mass state with a 
$q\bar q$ component in an octet
configuration. The two mesons $f_0(1710)$ and $f_0(1770)$ deviate from this
conjecture. It is possible that the small mass difference between these two states
leads to this unexpected mixing. Note that in the case of $f_0(2020)$ and $f_0(2100)$,
only the nominal masses are very close. The measured mass difference is 150\,MeV.

More important are the glueball contributions to the mesonic wave functions.
The probability that the glueball mixes into one of these resonances is \\[-1.5ex]

\renewcommand{\arraystretch}{1.2}
\hspace{-7mm}\begin{tabular}{cccccc} 
$f_0(1370)$\hspace{-3mm}&\hspace{-3mm}$f_0(1500)$\hspace{-3mm}&\hspace{-3mm}$f_0(1710)$\hspace{-3mm}&\hspace{-3mm}$f_0(1770)$\hspace{-3mm}&\hspace{-3mm}$f_0(2020)$\hspace{-3mm}&\hspace{-3mm}$f_0(2100)$\\
(5\er4)\%\hspace{-3mm}&\hspace{-3mm}$<5$\%\hspace{-3mm}&\hspace{-3mm}(12\er6)\%\hspace{-3mm}&\hspace{-3mm}(25\er10)\%\hspace{-3mm}&\hspace{-3mm}(16\er9)\%\hspace{-3mm}&\hspace{-3mm}(17\er8)\% \\
\end{tabular}
\renewcommand{\arraystretch}{1.0}

\begin{figure}[h]
\includegraphics[width=0.46\textwidth]{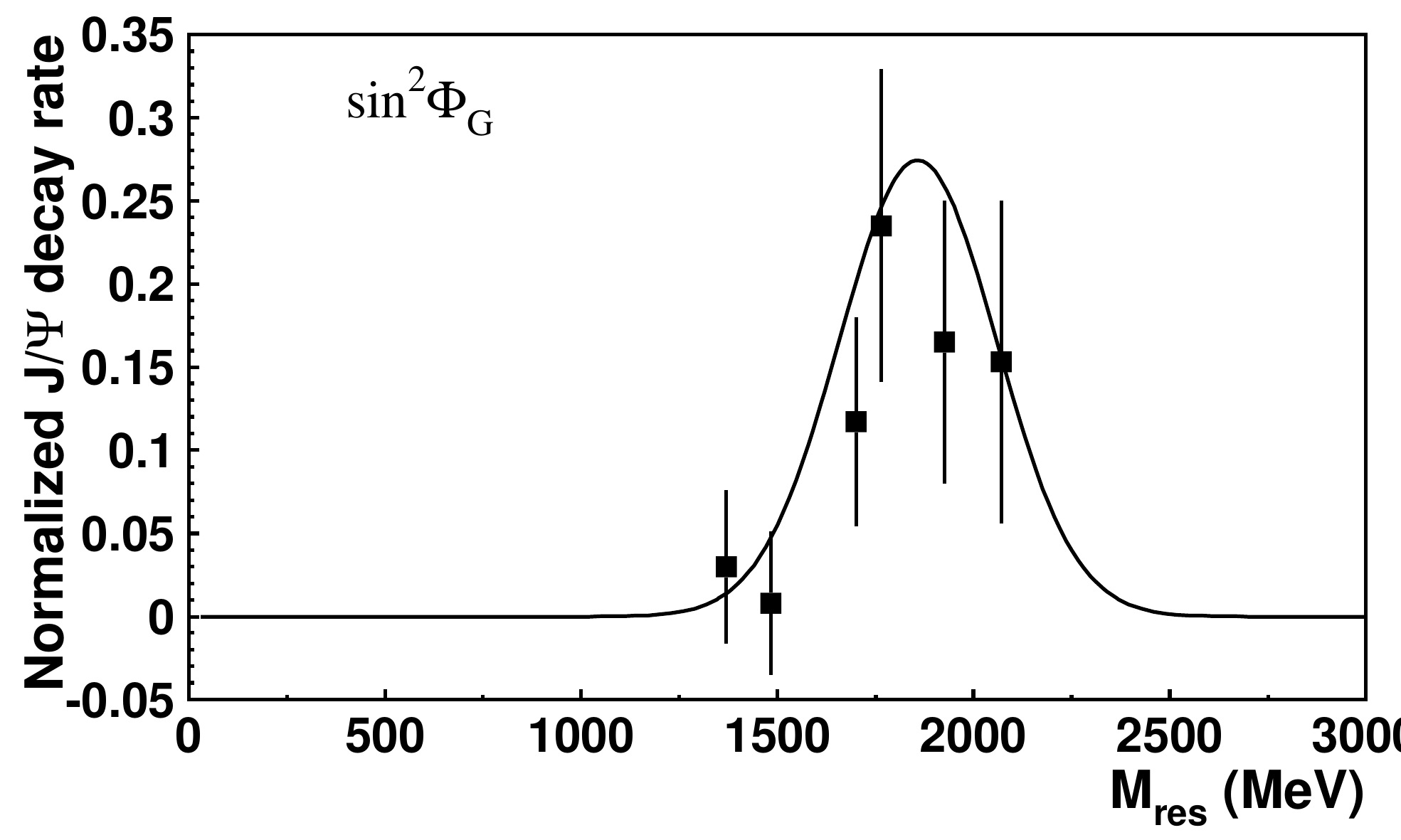}\vspace{-3mm}
\caption{\label{fig:glue}The glueball content of scalar mesons.\vspace{-3mm}
}
\end{figure}

The glueball is distributed, the sum of the fractional contribution is (78\er18)\%. 
A small further contribution (of about 10\%) can be expected from the two higher mass states 
$f_0(2200)$ and $f_0(2330)$. Figure~\ref{fig:glue} shows the fractional contribution of 
the scalar mesons to the glueball. The solid curve is a Breit-Wigner function with
mass and width $M=1865$\,MeV, $\Gamma=370$\,MeV, the area is normalized to one. 
The fractional glueball contributions to scalar mesons determined from the decays of scalar
mesons are compatible with the Breit-Wigner resonance
observed in production of scalar mesons in radiative $J/\psi$ decays. This is a remarkable
verification of the interpretation of the bump observed in radiative $J/\psi$ decays as
the scalar glueball of lowest mass.

The pair of resonances $f_0(1710)$ and $f_0(1770)$ was mostly interpreted as one
single resonance and was often identified with {\it the} scalar glueball~\cite{Sexton:1995kd}. 
The discovery of two scalar mesons, $f_0(1370)$ and $f_0(1500)$ (see \cite{Amsler:1995gf}
and references therein) stimulated Amsler and Close to propose a mixing scheme
where two scalar $q\bar q$ mesons and the scalar glueball mix to create the
three observed states $f_0(1370)$, $f_0(1500)$, $f_0(1710)$ ~\cite{Amsler:1995tu,Amsler:1995td}.
This paper led to a large number of follow-up studies with different mixing 
schemes~\cite{Burakovsky:1998zg,Lee:1999kv,Li:2000cj,Close:2001ga,Giacosa:2004ug,%
Close:2005vf,Giacosa:2005zt,Cheng:2006hu,Chen:2009zzs,Gui:2012gx,Janowski:2014ppa,%
Cheng:2015iaa,Frere:2015xxa,Vento:2015yja,Noshad:2018afw,Guo:2020akt}. These
all have one property in common: they impose that the sum of the fractional glueball 
contributions from these three resonances adds up to one. This we do not impose. We find that the sum over
six scalar resonances is (78\er18)\% and expect further 10\% from higher-mass
resonances. Within errors, the full glueball is covered.

A second important difference is the nature of the glueball. In earlier papers, the
glueball is seen as an intruder, as additional resonance entering the spectrum
of scalar $q\bar q$ mesons (with possibly further tetraquarks or hadronic molecules).
Glueball and scalar mesons mix but there is supernumerary of scalar states. 
We see the glueball as enhancement in the yields of ordinary scalar mesons. The decays
of these scalar mesons show that their wave functions must contain  a fraction of
the glueball. But the glueball is not an additional meson, the glueball
shows up only as fractional contribution to the wave functions of scalar mesons. 

Summarizing, we have studied the decays of scalar iso\-scalar mesons. The decay
couplings were fit with the assumption that their wave functions contain three
components: $n\bar n$, $s\bar s$ and a glueball component. The $\pi\pi$ and $K\bar K$
decay modes are well described, the $\eta\eta$ decay mode only partly, the $\eta\eta'$ decay
mode is often at variance with the prediction. Since the glueball content is mostly determined  
by the $\pi\pi$ and $K\bar K$ decay modes, we extract the glueball component. It follows
that the glueball is spread over a large number of resonances. The sum of all fractional
contributions is close to one. The scalar glueball is thus not only identified in 
radiative decays of $J/\psi$ mesons but also by its fractional contributions to the wave
functions of the scalar isoscalar mesons that were produced in radiative decays of $J/\psi$ mesons. 

The masses of all eight scalar-isoscalar H and L states above 1\,GeV discussed here fall onto two regular 
trajectories in a $M^2, n$ plot (see Fig.~3 in~\cite{Sarantsev:2021ein}). None of 
them seems irregular. Apparently, the scalar glueball does not enter the spectrum of scalar mesons 
as supernumerous state. It seems not to decay directly into two mesons but to decay only by
mixing with regular scalar isoscalar mesons.

\section*{Acknowledgement}
{\it Funded by the NSFC and the Deutsche Forschungsgemeinschaft (DFG, German 
Research Foundation) through the funds provided to the Sino-German Collaborative
Research Center TRR110 “Symmetries and the Emergence of Structure in QCD”
(NSFC Grant No. 12070131001, DFG Project-ID 196253076 - TRR 110) and 
the Russian Science Foundation (RSF 16-12-10267).}

\end{document}